# Yielding and flow in aggregated particulate suspensions

Peter J. Scales[1*], Shane P. Usher[1], Maria Barmar Larsen[2], Anthony D. Stickland[1], Hui-En Teo[1], Ross G. de Kretser[1] and Richard Buscall[3]

1 Centre of Excellence for Enabling Eco-Beneficiation of Minerals, Department of Chemical Engineering, University of Melbourne, 3010, Victoria, Australia
2 Department of Chemistry and Bioscience, University of Aalborg, Aalborg, Denmark
3 34 Maritime Court, Exeter, EX2 8GP, UK

* Author for correspondence: email: peterjs@unimelb.edu.au

## Abstract

The flow and consolidation of strongly flocculated particulate suspensions in water is common to a range of processing scenarios in the minerals, food, water and wastewater industries. Understanding the compressive strength or resistance to consolidation of these suspensions is relevant to processes such as filtration, centrifugation and gravity settling, where the compressive strength defines an upper boundary for processing. New data for the compressive strength of consolidating flocculated particulate suspensions in water, including alumina and calcium carbonate, are compared with earlier data from the literature and from our own laboratories for several systems, including two earlier sets of data for alumina. The three sets of data for the compressive strength of alumina agree well. Differences are noted for data measured in shear between our own laboratories and others. New data for the shear strength of AKP 30 alumina are also presented and, although the agreement is not as good, the difference is implied to be due to wall slip associated with a difference in measurement techniques.

A simple non-linear poro-elastic model of the compressive strength was applied to the eight sets of compressive strength data and was found to account for most features of the observed behaviour. The agreement strongly supports the mechanistic failure mode in compression for these systems to be one of simple strain hardening. The one feature that it does not account for without invoking a 'ratchet' is the irreversibility of consolidation. It is however suspected that wall adhesion might provide such a ratchet in reality, since wall adhesion has been neglected in the analysis of raw compressive strength until recently, notwithstanding the pioneering work of Michaels and Bolger (1962).

Overall, the data analysis and fitting presented herein indicate a new future for the characterisation of aggregated particulate suspensions in shear and compression whereby a limited data set in both compression and shear, albeit targeted across a wide concentration range, can now be used to predict comprehensive curves for the shear yield stress and compressive yield stress of samples using a simple poro-elastic model. The voracity of the approach is indicated through a knowledge that the behaviour of both parameters is scalar across a wide range of materials and across a wide range of states of aggregation.

# 1 Introduction

Strongly flocculated particulate suspensions in water are typical of industries from mining to wastewater processing. Above a critical solids concentration (denoted as $\phi_g$), these suspensions show an activation barrier to the initiation of flow (yielding behaviour) in both shear and compression. At increasing solids concentrations, the force to initiate flow behaviour rises in an exponential fashion [1]. The measurement of yielding in shear ($\sigma_y$) was simplified by Nguyen and Boger [2, 3] with the introduction of the vane shear yield stress method, although a range of authors had previously reported apparent or Bingham yield stress values in flocculated suspensions through extrapolation of shear stress-shear rate rheological data to zero shear rate. The latter are not considered here since the study of yielding is, at a philosophical level at least, a static to dynamic transition rather than the opposite.

Measurement of a peak stress in shear during constant rate flow start-up has been observed across a wide range of particulate suspensions with systematic changes noted as a function of solids concentration and other parameters that are known to change the strength of inter-particle attraction, for example: pH, salt concentration or the addition of simple molecular additives [1, 4-9]. Phenomenological modelling of these same systems demonstrated that the changes in shear yield stress as a function of parameters such as pH, salt concentration and even ion type [10, 11] are scalar as a function of solids concentration. This indicates that the mechanism of failure (yielding) and initiation of unsteady flow in these systems is cooperative. Some authors hypothesized that this scalar correlation to the strength of adhesion between particles implied that the measured shear yield stress was a pseudo material property, although such an analogy was treated with a useful level of scepticism [12]. None the less, most researchers and practitioners in particulate processing were happy to accept that its measurement was useful for a range of process control scenarios relating to flow, mixing and pipeline start-up, with empirical measurements common [13].

More recent research of the mechanism of yielding in shear has shown that the process can be described well by an analogy to cage melting [14] whereby under an applied stress, the particulate network first strain hardens followed by interparticle bond breakage. At low solids concentrations the system then yields and flows, with the strain at the point of yield consistent with the interparticle interaction length. However, the more usual observation for systems with a concentration much greater than $\phi_g$ is that after bond breakage, the network undergoes Brownian reformation that is in competition with the rate of shear and hence, we observe shear dependent strain softening leading to cage melting (yielding) [15-17]. At higher strains, we then see unsteady-steady flow. The strain at yield is typically of order 1, consistent with the cage melting analogy, and the measured yield stress is dependent on the rate of measurement. Aiding the previous debate [12], the more detailed analysis demonstrates that the measured shear yield stress is not a singular parameter for flocculated suspensions. Indeed, Nguyen and Boger [2] demonstrated this behaviour in their early work and chose a rotation rate for their vane based shear yield stress measurements based on the minimum in the observed peak stress.

Systematic data for yielding in compression are not as widely reported although work by Channell and Zukoski [18], Green et al. [19] and the extensive work by Zhou et al. [20, 21] indicates a systematic scalar relationship between the peak stress measured in shear ($\sigma_y$) and the pressure required to achieve the same solids concentration at equilibrium through uniaxial compression. The latter data was typically achieved through filtration tests to equilibrium although data closer to the gel point are sometimes measured through either sedimentation [22, 23] or centrifugal methods [24].

As indicated, these flocculated (strongly cohesive) particulate suspensions undergo differential compression when they consolidate under gravity, or in a centrifuge, or when they are pressure-filtered or dried, whereby the particulate network reduces in volume as liquid is expelled. Since the particles are sticky and a finite volumetric strain cannot sensibly be distinguished from a change in volume fraction, the volumetric Hencky strain ($\varepsilon_H$) is taken to be given by [25]:

$$\delta\varepsilon_H = \delta \ln\phi \qquad [1]$$

and the bulk modulus by:

$$K(\phi) = \mathrm{d}P(\phi)/\mathrm{d}\ln\phi \qquad [2]$$

A simple prescription for the compressive strength ($P$) then follows, thus:

$$P(\phi) = \int_{\phi_0}^{\phi} K(x)dln(x) = a\int_{\phi_0}^{\phi} G(x)dln(x) \qquad [3]$$

where $G$ is the shear modulus and the order-one constant $\alpha$ is equal to 5/3 for the case of central forces only (by way of illustration). It follows from equation 3 that, strictly, $P = P(\phi, \phi_0)$, although the dependence on the starting concentration ($\phi_0$) is predicted to be weak when the modulus increases rapidly with concentration, as it does in practice (typically $\phi^{4.5\pm0.5}$) [18, 21, 25-27]. Hence, it is usually found that $P \approx P(\phi)$, i.e., the compressive strength looks like a pseudo-material property, also referred to as the compressive yield stress.

It is usually found also that compression is largely irreversible (there is sometimes some minor recovery), which is at odds with the above poro-elastic prescription, from which one would expect 20% or more recoverable strain, depending upon volume fraction, given the exponent of ~4. The particles are however adhesive and stick to the walls of the vessel, so a 'ratchet' [28, 29] can be invoked to explain this feature. It has been found also that above the gel-point a critical level of pressure needs to be exceeded before anything happens [25]; this being why $P(\phi)$ is often called the compressional yield stress. This feature certainly is at odds with the relationship above. There is however a reluctance to abandon Equation 3 nevertheless, because it does account for one very significant and widespread finding, viz. that $p \sim K \sim G \gg \sigma_y$ (away from the gel-point). Flocculated particulate suspensions are however usually adhesive as well as cohesive and wall adhesion can explain this anomaly on laboratory scales, since yield at the walls is a necessary precursor to compression [30, 31].

The adhesion can be characterised by a wall shear stress $\sigma_w$ [30], related to the bulk yield stress by,

$$\sigma_w \le \sigma_y \equiv \gamma_c G, \qquad [4]$$

where $\gamma_c$ is the apparent yield strain, defined as the ratio of yield stress to linear modulus. This definition adds nothing as such, but it provides a very convenient way of parameterizing adhesive strength [31, 32].

Wall adhesive effects are expected to be unimportant when [30, 32]:

$$\frac{20L\gamma_c G(\phi)}{3wP(\phi)} \ll 1, \qquad [5]$$

where $w$ and $L$ are the horizontal width and vertical length of the sample, but significant otherwise. It is also easy to show from equations 3 and 4 that wall effects are *always* expected to be important when and where the volume-fraction is very close to either the starting concentration, $\phi_0$, or the gel-point $\phi_g$, whichever is the greater, almost regardless of the value of $\gamma_c$.

## 2 Experimental Section

This work looks at historical published data as well as some new laboratory data. The historical data was taken from various publications and includes work on attapulgite clay, silica and latex (Buscall et al., [33]), alumina (Channell and Zukoski [18], Zhou et al. [21]), and kaolin (Aziz et al. [34]). The new data was generated with two types of materials, namely AKP 30 alumina (labelled Kristjansson herein) and calcium carbonate (labelled HE Teo herein). The alumina was an aluminium oxide (99.99%, APK-30, Sumitomo Shoji Chemicals Co., Tokyo, Japan). The 360 nm alumina particles were suspended in a $10^{-2}$ M $KNO_3$ solution and sonicated at 200 Watt (Branson Sonifier 450) for 2 minutes to break up lumps, with the pH adjusted to 5. After further sonication for 1 minute the suspension was then allowed to rest for 24 hours. The calcium carbonate was an industrial sample from Omya, Australia (Omyacarb 2) and was dispersed as for the alumina with a sonication horn but at pH 11. Prior to measurement, the suspension was coagulated at its isoelectric point (pH 8.0±0.1) and allowed to equilibrate for 2 hours. The Sauter mean diameter of the calcium carbonate was 4.5 µm (Malvern Mastersizer 2000).

The AKP 30 alumina was aggregated in two ways. The first involved coagulation to the isoelectric point of the suspension. In this case, the pH of the suspension was adjusted to 9.0 ± 0.1 and allowed to rest for two hours prior to measurement. Different volume fractions were obtained by diluting a stock suspension with 0.01 M $KNO_3$ solution and adjusting the pH. The volume fraction was determined by weight loss on drying, where the samples were left to dry for 24 hours at 100°C (Lab-line Due-Vac oven, Melrose Park, ILL).

In the second method of aggregation, suspensions were flocculated with polyacrylic acid and a polyacrylamide polymer. Alumina suspensions with an electrolyte concentration of $10^{-3}$ M $KNO_3$ were prepared as described previously. A stock particulate suspension was diluted to 2.5 w/w% with $10^{-3}$ M $KNO_3$ solution. The diluted suspensions were then sonicated for 1-2 min and the pH adjusted to 7.29 ± 0.1. A polyacrylic acid solution (MW = 250,000 g/mol, Aldrich Chemical Company, USA), was prepared at 0.1 w/w% in demineralised water. The solution was stirred for 1 hour, to ensure complete dissolution of the polymer, using a magnetic stirrer. A non-ionic polyacrylamide solution (0.2 w/w% Magnafloc LT20) was produced by dissolving the dry polymer in ethanol and mixing it with demineralised water (0.2 g powder to 2 mL ethanol in a 100 mL solution). The covered mixture (to avoid UV degradation of the polymer) was shaken overnight. An hour before flocculation, a 0.01 w/w% solution was produced by diluting and stirring for 1 hour.

The suspensions were flocculated in a 1000 ml baffle reactor as described by Holland and Chapman [35]. After 5 minutes mixing at 500 rpm (Heidolph, RZR 2020 control), the speed was reduced to 330 rpm and the PAA solution was added and mixed for 1 minute. After addition of PAA, the Magnafloc LT20 solution was added to the suspension and mixed for 20 seconds. The suspension was left for the flocs to settle, and the settling rate measured. After the flocs had settled, the stirrer and the baffles were removed from the suspension and the supernatant was decanted.

The shear yield stress was measured with a vane on a Haake Viscometer (HAAKE VT550, Kahlsruhe, Germany). A range of vanes were utilised to cover a wide range of peak stresses. The different vane dimensions employed are shown in Table 1. The yield stress measurement was performed by loading the sample in a suitable container and for the sample coagulated to the iso-electric point and then shearing it well to avoid potential thixotropic effects that would influence the results. The vane was lowered into the suspension until the suspension covered the vane, plus a further distance corresponding to the vane radius. The sample was allowed to rest for 2 minutes. For the polymer flocculated sample, the suspension was not mixed after loading. A shear rotation rate of 0.2 rpm was then applied to the vane and the yield stress was calculated according to the method of Nguyen and Boger [2]. The Haake had a limiting torque of approximately 30,000 μNm corresponding to a maximum limiting stress of approximately 10 kPa (using the smallest vane).

Table 1: The vane dimensions employed in shear yield stress measurements.

| Vane | Height $H$ (mm) | Diameter $D_v$ (mm) |
|---|---|---|
| 1 | 15.145 | 9.96 |
| 2 | 20.20 | 20.21 |
| 3 | 30.17 | 25.02 |
| 4 | 50.0 | 25.0 |
| 5 | 75.0 | 25.0 |
| 6 | 100 | 50 |
| 7 | 200 | 100 |

Compressive yield stress measurements were completed using a combination of sedimentation, filtration [36, 37] and centrifugation tests [38]. Filtration tests were conducted on a piston driven filtration rig that used a linear encoder to monitor piston height. The device and its operation are described elsewhere [36], inclusive of the measurement of the final volume fraction of suspensions for a given equilibrium filtration pressure. This value was compared to the final volume fraction as determined by weight loss on drying and the agreement was found to be good.

For centrifugation, the method used polycarbonate centrifuge tubes (Naglene) with a height of 103 mm and inside diameter of 26.5 mm. A flat base inside the tube was formed by adding a small amount of epoxy resin into each tube. The measurement was then performed by measuring the initial height of the tubes. As the epoxy resin was not completely even, the height was determined with a steel ruler (precision 0.1 mm, Digital Caliper) from an average value of six measurements round the tube. Sample heights between 7-50 mm were employed. The suspensions were tested at 20°C. The samples were run until equilibrium was obtained (i.e. a constant bed height). The sample was divided into about 10 slices. Each sub-sample was removed by scraping the sample from the tube using a flat spatula fixed at the desired sample height and the volume fraction for the sub-sample was found by weight loss on drying. Compressive yield stress data was calculated according to the method described by Green and Boger [39].

In sedimentation tests, an initial concentration below the gel point was chosen. Glass measuring cylinders with different initial suspension heights were left to settle and the sediment-liquid interface measured until equilibrium was reached. The final height was then measured and the resulting volume fraction and the compressive yield stress calculated using the final average volume fraction and compressive stress as described in [40].

## 3 Results and Discussion

Figure 1a shows a compilation of compressive strength data for some coagulated systems, including extensive data for colloidal alumina from 2 different labs and 4 different workers. Most of the suspensions depicted in Figure 1 comprise either spheroidal or tuberoidal particles. The particle size range covered is enormous (from 4.5 μm $CaCO_3$ down to 26 nm $SiO_2$), although data for fine kaolin (lozenges) and attapulgite (long needles) is also shown. The data are scaled on the ordinate in Figure 1b whereby the data has been shifted vertically such that all data match a strength of 20 Pa at a volume fraction of 0.1. Although an arbitrary reference point in terms of the shift, a master curve results, implying the mechanism of failure to be both consistent and cooperative across a wide range of aggregated colloidal particulates. The only significant departure from the master curve is for $SiO_2$, which shows an obvious gel-point. The $SiO_2$ gels were however very strong by virtue of their small particle size (26 nm diameter) and it was only with this material that the methods used (centrifugation or pressure filtration) could measure near the gel-point. The gel-point of the colloidal alumina could well be similar to that of the silica though, as it looks to be 0.05 or less; see also Figure 2, which shows shear yield stress data for the colloidal alumina (this work).

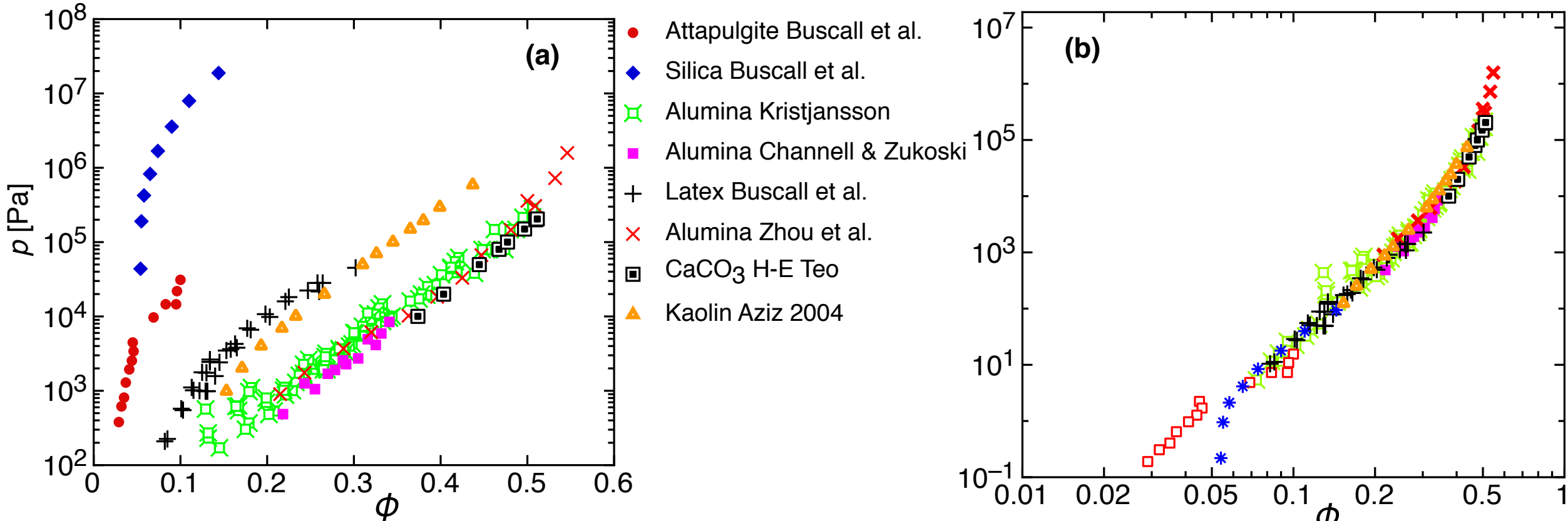

Figure 1: Plots, semi-log and log, of compressive strength versus volume-fraction for a range of inorganic particulate suspensions together with polystyrene latex. The new (unpublished) data for AKP-30 alumina and calcium carbonate are those labelled 'Alumina Kristjansson' and '$CaCO_3$ HE Teo.' In fig. 1(b) the data in 1(a) have been scaled by shifting data on the ordinate such that all data goes through a fixed point of 20 Pa at a volume fraction of 0.1.

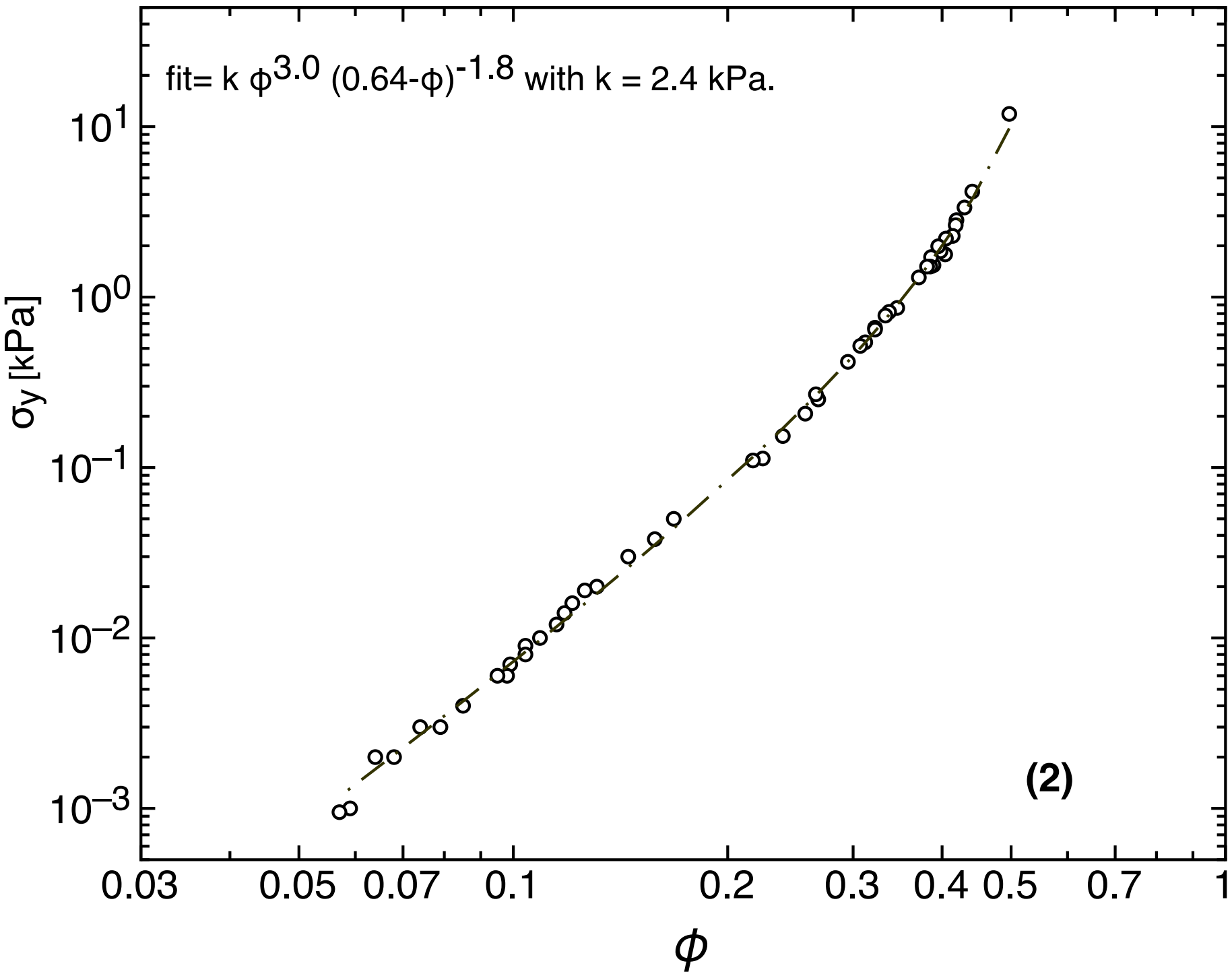

Figure 2: Plot of shear yield stress data (this work) versus volume-fraction for AKP-30 alumina, fitted as shown.

The concentration dependence of the shear yield stress is weaker than that of the compressive yield strength in the power-law region. The power-law of 3.0 is typical for shear, whereas the shear storage modulus is found to show a power of $> 4$ [18, 21, 25, 27]. This tells us that the apparent critical strain defined by the right-hand side of equation 4 increases with concentration. If equation 3 is correct then, it should be possible to predict the compressive strength from the shear strength using equations 3 and 4. The blue and red lines in Figure 3 show the results.

The blue line was obtained by supposing that the apparent critical shear strain is constant (and 0.005), which we know it is not, and the red line by assuming that it is proportional to $\phi$. The agreement between the red line experimental data is excellent for $\phi < 0.4$. Above 0.4, it is not, although there are good reasons why there might be a difference there, which space does not permit discussing here, except to say that neglect of the osmotic pressure of the particles and the potential for the failure to be dominated by sliding friction at higher solids concentrations are two aspects [41]. The agreement below $\phi < 0.4$ is however, most gratifying.

Direct experimental evidence in favour of the model encoded in equations 3 and 4 is shown in Figure 4, where the experimental ratio of shear to compressive strength is plotted against $\phi$, together with the experimental uncertainty. The curves imply that the ratio starts at unity and decays rapidly with increasing volume fraction. This is exactly what equations 3 and 4 predict must happen by embodying the idea that particulate gels are strain-hardening in compression and in shear, show strain hardening with failure at strains equivalent to the length scale of the interparticle force at low volume fractions but shift to a cage melting behaviour dominated by strain softening in shear as the volume fraction increases. It alone is powerful evidence that equations 3 and 4, simple though they are, capture most of the observed behaviour and that failure in compression is mechanistically simpler than failure in shear, at least at volume fractions away from the gel point. It should be noted though that the compressional data for alumina have not been corrected for wall adhesion, which must contaminate them to some extent at the lowest concentrations [32], doing so, however, can only make the rise towards unity at low concentration in Figure 4 even steeper, reinforcing the point.

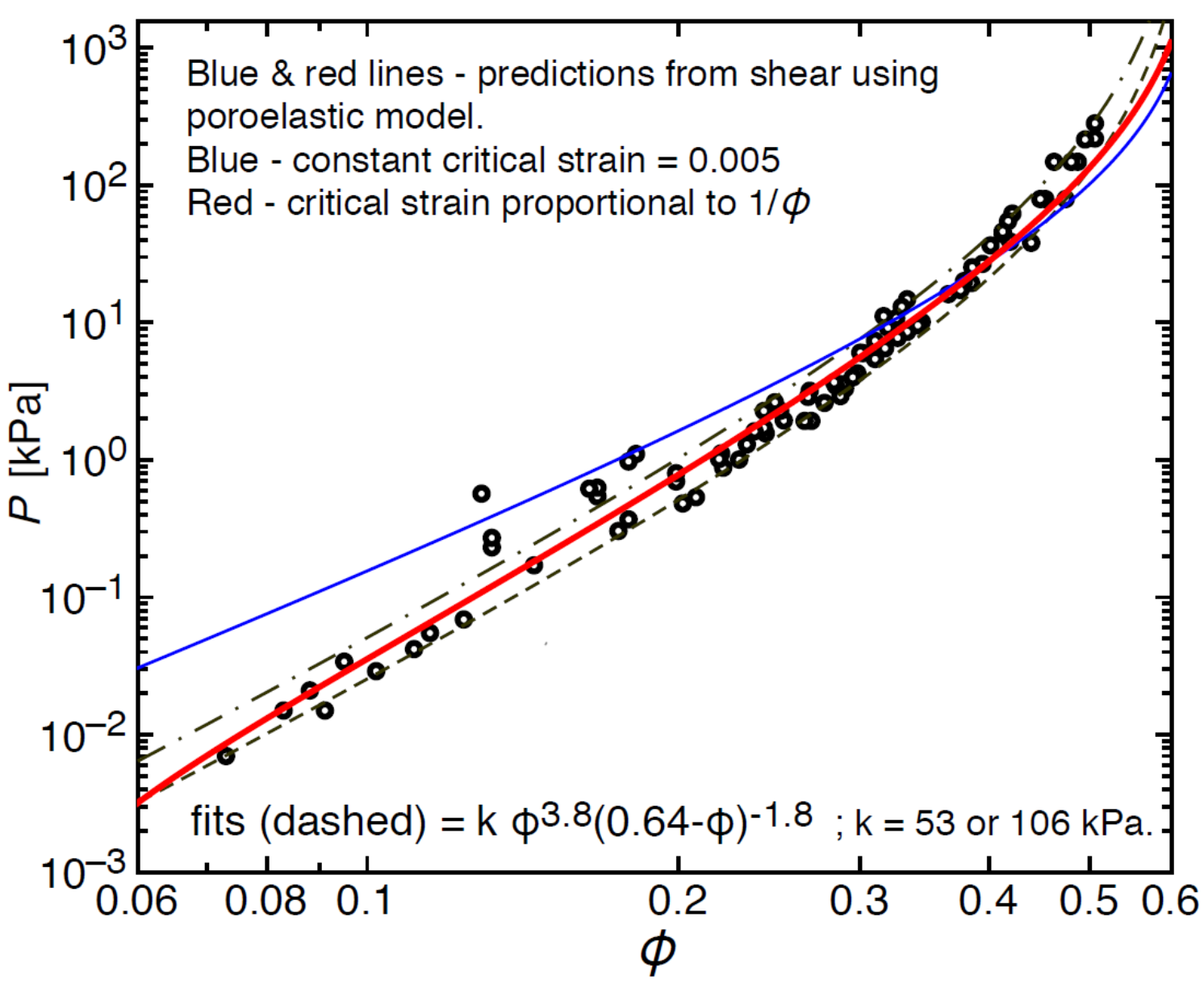

Figure 3: Compressional stress for AKP-30 Alumina, fitted as shown and compared with predictions made from the data in fig. 2 using equations 3 and 4 (see text).

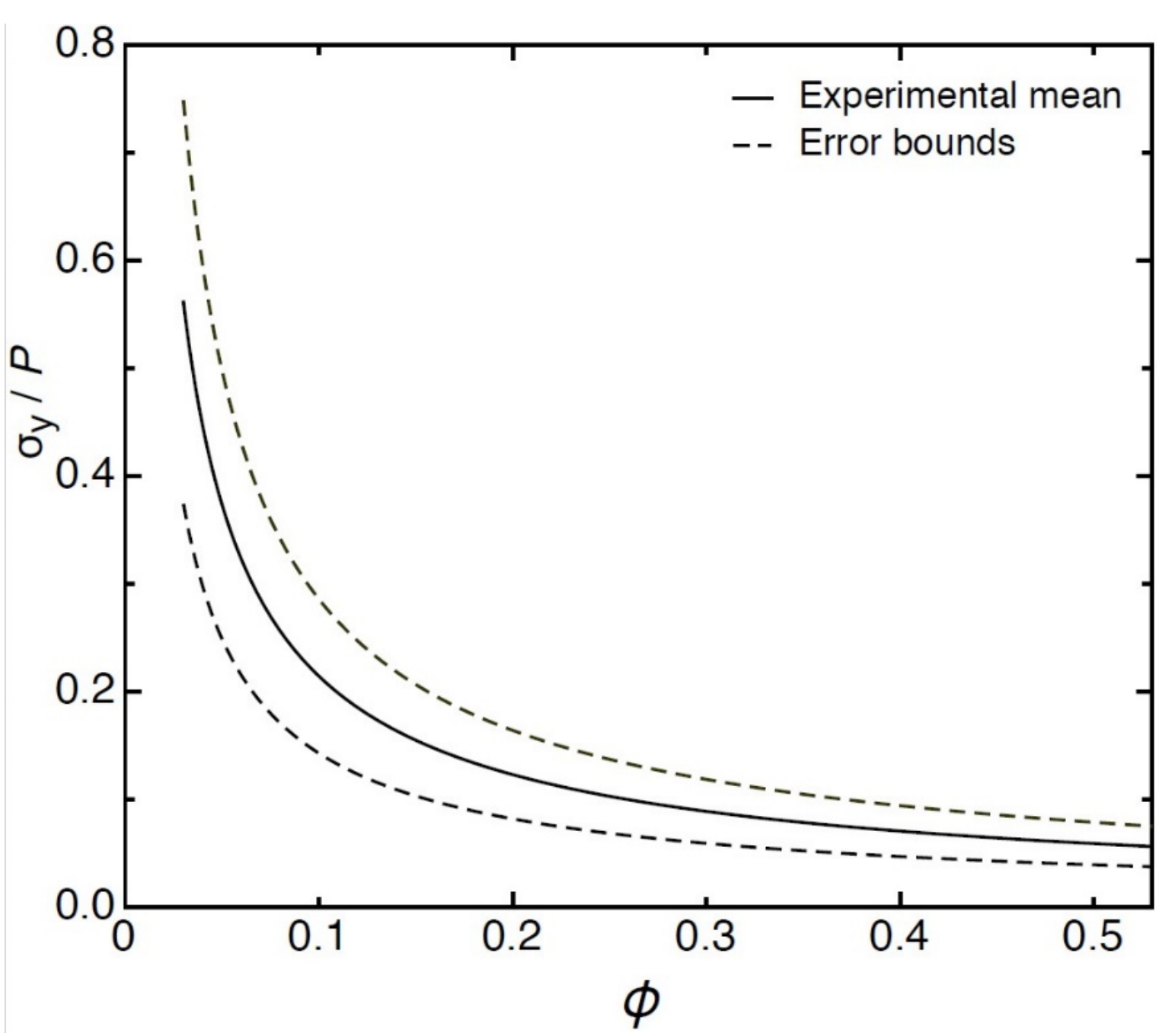

Figure 4: Plot of shear to compressional strength ratio obtained from the fits shown in Figures 2 and 3.

Whereas Figure 4 contains direct evidence in support of the simple constitutive model embodied in equations 3 and 4, it might be thought that the comparison with predictions made using these equations amounts to little more than curve-fitting. This would certainly be true were the choice of the form of the apparent critical strain function to be unconstrained and allowed to float. The argument then, lies in having independent evidence in favour of a roughly linear increase of apparent critical strain with volume fraction. That argument is as follows: that it has been widely observed [21, 25, 27, 42] that the exponents of the concentration dependence of the shear yield stress is typically found to be ~ 3 and the shear modulus typically ~4, i.e. they differ by ca. one. The weakness of this argument is that, although we see an exponent of 3 for the yield stress, we do not have a similar comprehensive set of data for the modulus of the alumina itself. Furthermore, we have ignored the shear data of Channell and Zukoski [18] who unusually, see the same exponent of ~5 for both properties. Not only is the exponent for the shear yield stress very different from ours, the magnitudes are overall dissimilar too, even though the compressive data agree. Their shear data, compared with ours, is shown in Figure 5.

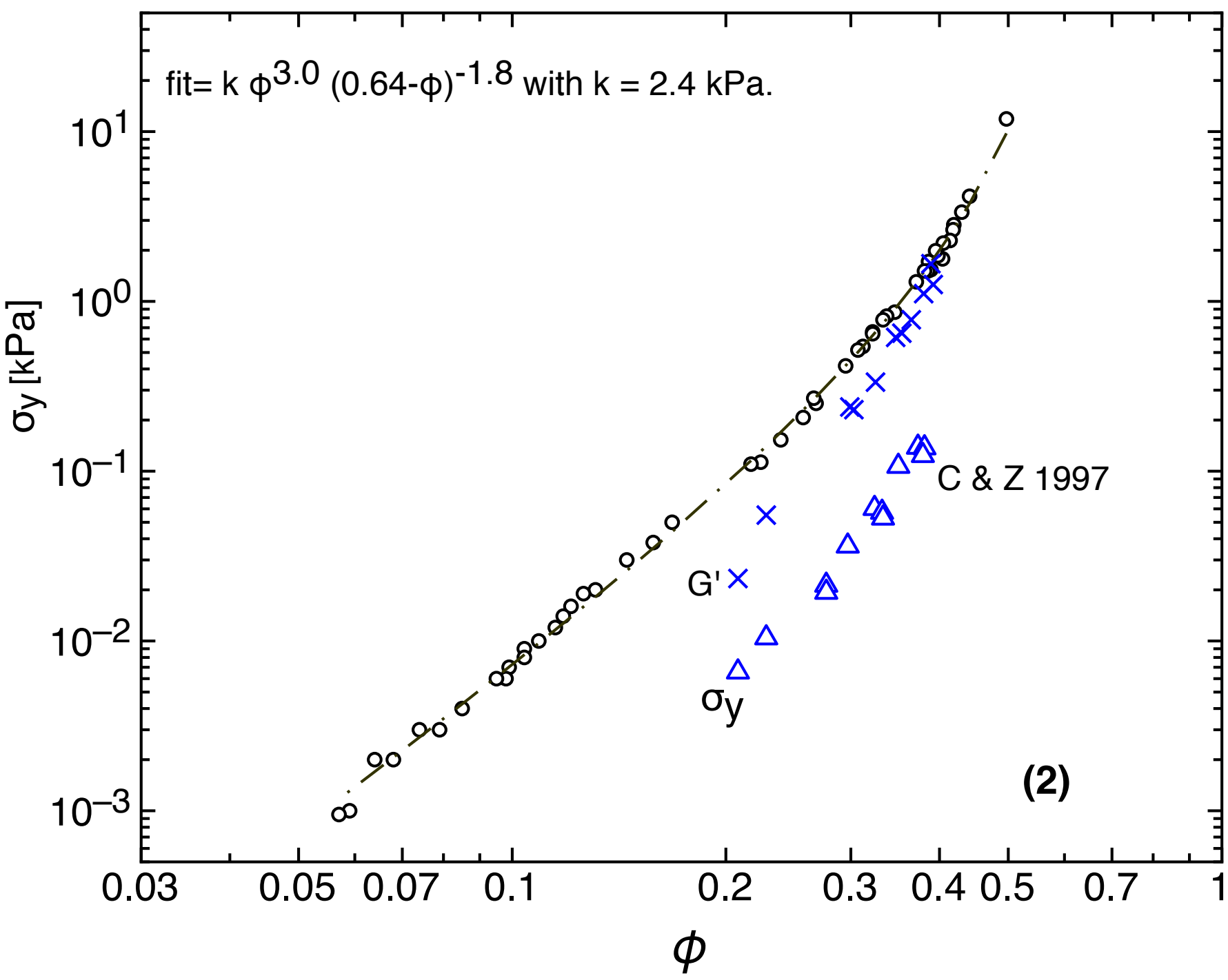

Figure 5: Comparison of the shear yield stress data (this work) with that of Channell and Zukoski [18]. Their G' data is also shown.

It can be seen that not only are yield stresses in the work of Channell and Zukoski about an order of magnitude smaller on the average, even their shear modulus is smaller than our yield stress. The high exponent and low magnitudes are causes for concern. Unlike us though, Channell and Zukoski used smooth concentric cylinders with small gaps to perform their measurements, hence a likely explanation of the difference is that they were seeing premature wall yielding and slip [43]. Another possibility is that their material might possibly have undergone the type of shear induced densification or granulation described by Firth [44] and Mills et al. [42], who showed that shear flow may or may not cause irreversible changes in microstructure, depending upon the detailed nature of the shear-rate history. Having said that, the samples herein are stirred prior to measurement and this is an unlikely explanation. We suspect that slip probably suffices to explain the difference, not least because it is known that materials of this type will slip at the outer cylinder, even with vane rotors, should the gap not be wide enough to prevent it [43]. It behoves us then, to repeat the measurements of yield stress and G' using both types of rheometer tool on the same batch of alumina in order to prove the point.

Were the material of construction of our centrifuge tubes to have been the same as that of Channell and Zukoski's [18] rheometer tools then we could perhaps have used their shear data, or an extrapolation of it to lower ϕ, to correct figures 3 and 4 for wall adhesion using the methods described by Lester and Buscall [32, 38]. We would however have found the error to be insignificant. Until such time as shear data for glass and polycarbonate become available, all that we can do in the first instance is to calculate an upper bound correction from the true or cohesive yield stress. This correction is shown by the continuous blue line labelled C1 in Figure 6a. C1 assumes that $\sigma_w = \sigma_y$ (i.e. $\lambda$=1)

We know that C1 is an over-correction because the divergence of the corrected ratio then implies a gel-point of ca. 0.09, whereas the $\sigma_y$ data in figure 2 extend below that; they imply a gel-point of 0.05 or less perhaps. Thus, what one can then do is to write $\sigma_w = \lambda\sigma_y$ and then find the value of $\lambda$ by shooting that brings the divergence down to 0.05. In this case a value of just less than 1/6 is needed in order to cause the strength ratio to diverge at 0.05. The second plot shows the original compressive strength data corrected using $\lambda$ = 1/6. The correction is not that large because fairly wide centrifuge tubes (26.5 mm) were used in this case. Such corrections become much more serious at, say, 10 mm or less. It does however highlight that the use of narrow centrifuge tubes could lead to the incorrect conclusion that there is no gel point in the system or that the gel point is lower than reality.

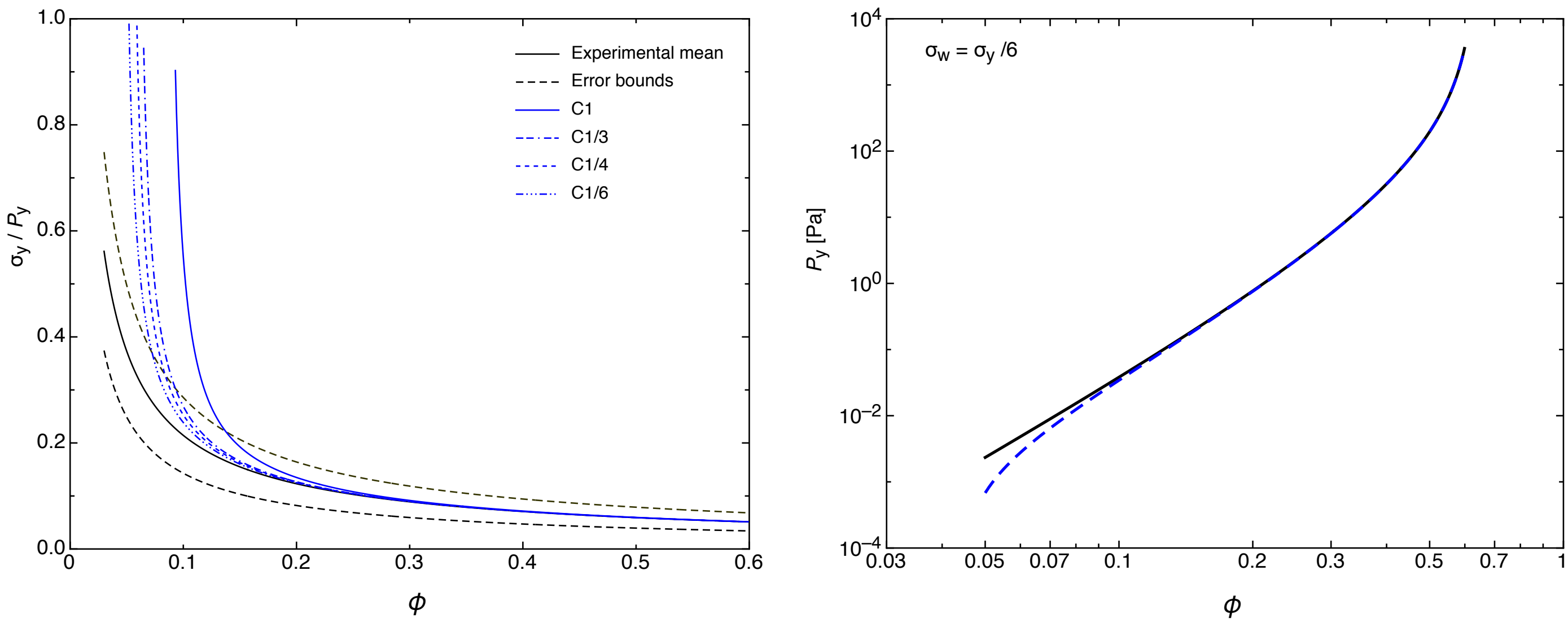

Figure 6a: Over-corrected strength ratios are shown in blue (see text for details) and

Figure 6b: Compressive strength data from figure 3 corrected using $\lambda = \sigma_w/\sigma_y$ = 1/6 where the solid line is the original data and the dashed line is corrected for wall adhesion.

It is interesting to note that the gel point of 0.05 or less implied by figure 2, compares well with that estimated from gravity batch settling in short tubes, this being 0.05 as well. It also highlights that both measurement and prediction of the shear and compressive yield stress of aggregated suspensions can be complimentary. Knowing that there is a simple method to predict the relationship between the shear and compressive yield stress and that both are scalar, measurement of full data sets as a function of solids concentration is now not a necessary condition to fully characterize a particulate suspension in shear and compression. Indeed, many laboratories are limited in the range of stresses that can be measured accurately using a vane tool on a simple rheometer and equally, measurement of the full compressibility curve for some materials, especially biological materials that show non-quadratic filtration behaviour and those with a very low gel point [45] require a range of techniques (sedimentation, centrifugation and filtration) to measure data across a wide range of volume fractions [46]. The analysis of the data is equally complicated [47]. Equally, it is easier for many systems to measure data close to the gel point through compression using sedimentation, at intermediate solids through shear, and at very high solids in compression using filtration. There now exists the possibility to exploit the measurement techniques over a particular solids range in both shear and compression that are easiest for a particular sample to produce a comprehensive characterization in both shear and compression.

## 4 Conclusions

The scaling of compressive strengh data for nominally spherical particles shown here strongly favour the simple 'ratchet elastic' constitutive model encoded in equations 3 and 4. The fact that both particle size [27] and now, apparently, shape can be scaled together in figure 1a is remarkable. That the compressive strength can be predicted from shear data provides further support for the model embodied in equations 3 and 4. That this simple model cannot account for irreversibility without hand-waving, nor for critical or yield-like behaviour, is of concern although wall adhesion may suffice to alleviate these concerns. It certainly looks as if it can in principle, but whether it does or not quantitatively is another matter, although work aimed at finding out is in progress. Inaccuracies in the data are only likely at concentrations close to the gel point and in measurements where the wall adhesion is expected to be a significant contributor to the total measured stress. Truncation of the analysis and appropriate choice of measurement apparatus can reduce the errors significantly. Overall, the data analysis and fitting presented herein indicate a new future for the characterisation of aggregated particulate suspensions in shear and compression whereby a limited data set in both compression and shear, albeit targeted across a wide concentration range, can now be used to predict comprehensive curves for the shear yield stress and compressive yield stress of samples using a simple poro-elastic model. The voracity of the approach is indicated through a knowledge that the behaviour of both parameters is scalar across a wide range of materials and across a wide range of states of aggregation.

## 5 Acknowledgements

The authors acknowledge the seminal work of PC Kapur in collecting and scaling a large amount of the original shear yield stress data from our laboratories and congratulate him on a wonderful career in science and engineering. The authors acknowledge the funding support from the Australian Research Council for the ARC Centre of Excellence for Enabling Eco-Efficient Beneficiation of Minerals, grant number CE200100009.